\documentstyle[11pt,newpasp,twoside,psfig,graphicx]{article}
\reversemarginpar

\def\etal{{et\,al.}}
\def\msun{M$_{\odot}$}
\def\rsun{R$_{\odot}$}
\def\mdot{$\dot M$}
\def\degs{\ifmmode ^{\circ}\else$^{\circ}$\fi}
\def\amin{\ifmmode ^{\prime}\else$^{\prime}$\fi}
\def\asec{\ifmmode ^{\prime\prime}\else$^{\prime\prime}$\fi}
\newbox\grsign \setbox\grsign=\hbox{$>$}
\newdimen\grdimen \grdimen=\ht\grsign
\newbox\laxbox \newbox\gaxbox
\setbox\gaxbox=\hbox{\raise.5ex\hbox{$>$}\llap
     {\lower.5ex\hbox{$\sim$}}}\ht1=\grdimen\dp1=0pt
\setbox\laxbox=\hbox{\raise.5ex\hbox{$<$}\llap
     {\lower.5ex\hbox{$\sim$}}}\ht2=\grdimen\dp2=0pt
\def\gax{$\mathrel{\copy\gaxbox}$}

\def\grs{GRS 1915+105}

\DeclareRobustCommand{\ion}[2]{%
\relax\ifmmode
\ifx\testbx\f@series
{\mathbf{#1\,\mathsc{#2}}}\else
{\mathrm{#1\,\mathsc{#2}}}\fi
\else\textup{#1\,{\mdseries\textsc{#2}}}%
\fi}

\begin{document}

\title{The binary components of GRS 1915+105}
 \author{Jochen Greiner}
\affil{Astrophysical Institute Potsdam, 14482 Potsdam, 
Germany}

\begin{abstract}
I summarize recent near-infrared spectroscopy of the microquasar GRS 1915+105 
with the ESO/VLT which allowed to (i) identify the donor and (ii) measure
the orbital period and radial velocity amplitude. Assuming that the jet
ejections occur perpendicular to the accretion disk and orbital plane, 
the mass of the black hole is determined to $M = 14 \pm 4$ \msun.
I discuss the implications of this mass determination on (a) the understanding
of the large-amplitude X-ray variability in GRS 1915+105,
(b) binary formation scenarios, and (c) models to explain the stable
QPO frequencies in GRS 1915+105 and GRO J1655-40.
\end{abstract}

\section{Introduction}

GRS 1915+105 (Castro-Tirado \etal\ 1994) is the prototypical microquasar, 
a galactic X-ray binary
ejecting plasma clouds at v$\approx$0.92\,c (Mirabel \& Rodriguez 1994).
It exhibits unique X-ray variability patterns (Greiner \etal\ 1996) 
which have been interpreted as
accretion disk instabilities leading to an infall of parts of the
inner accretion disk (Belloni \etal\ 1997).
Based on its X-ray properties GRS 1915+105 was suspected to be the 
most massive stellar black hole candidate in
the Galaxy (Morgan \etal\ 1997). Besides GRO J1655-40, GRS 1915+105
is one of only two galactic sources which are thought
to contain a maximally spinning black hole (Zhang \etal\ 1997).
It is therefore of great importance to know some details about the
system components in the GRS 1915+105 binary in order to understand 
the conditions which lead to
the unique X-ray, radio, and infrared characteristics.

\section{Observations and Results}

\subsection{The donor}

 \begin{figure}[th]
   \vbox{\psfig{figure=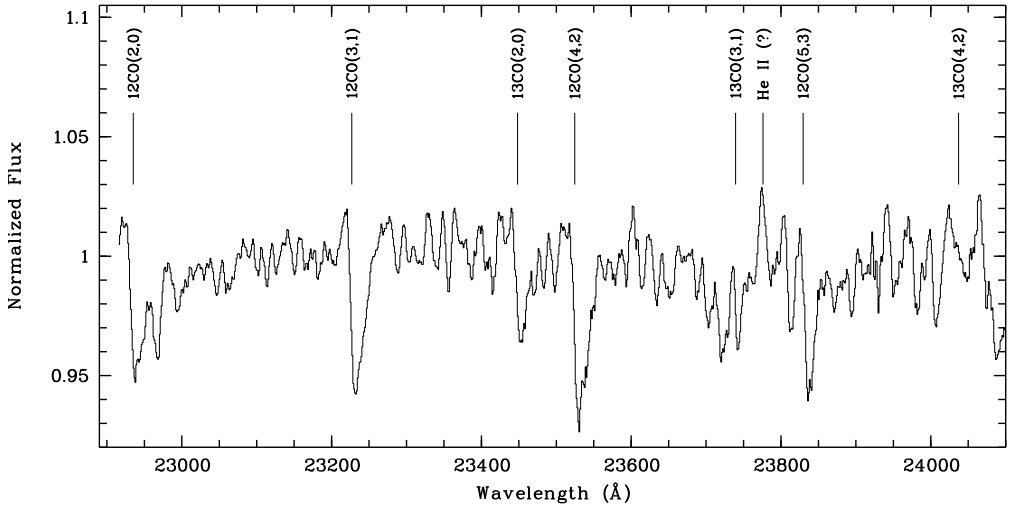,width=13.6cm,angle=270,%
           bbllx=4.7cm,bblly=3.cm,bburx=17.0cm,bbury=27.cm,clip=}}\par
   \vspace{-0.2cm}
    \caption[irsp]{Mean K band spectrum of GRS 1915+105.
     CO bandheads are clearly discovered and marked by vertical lines. 
       The presence of the $^{13}$CO isotope
       and the equivalent width ratio of $^{12}$CO to $^{13}$CO suggests a
       classification of the donor as a late-type giant.
      \label{veil}}
 \end{figure}

\grs\ is located in the galactic plane at a distance of $\sim$11 kpc 
(Fender \etal\ 1999) 
and suffers extreme extinction of 25--30 mag
in the visual band. We therefore obtained spectroscopic observations
in the near-infrared H and K bands 
using the ISAAC instrument on the VLT-Antu telescope.
Based on spectra taken in the 5 wavelengths bands 1.56--1.64 $\mu$m,
1.63--1.72 $\mu$m, 2.05--2.17 $\mu$m, 2.17--2.29 $\mu$m and 2.29--2.41 $\mu$m
we successfully identified absorption features
from the atmosphere of the companion (mass donating star) in the
GRS 1915+105 binary (Greiner \etal\ 2001a). The detection of $^{12}$CO and
$^{13}$CO band heads  (Fig. \ref{veil}) plus a few metallic absorption lines 
suggests a K-M spectral type and luminosity class III (giant).
Comparing the equivalent widths
of the $^{12}$CO(3,1) $^{12}$CO(4,2) and $^{12}$CO(5,3) bandheads
between the GRS 1915+105 spectrum and that of a K2\,III standard star
(observed with the same settings) implies that the donor of GRS 1915+105 
has an unveiled magnitude of 14.5--15.0 mag (prior to extinction correction).
This in turn implies an absolute magnitude of 
$M_{\rm K}$ = --2...--3, consistent with the giant classification.
This K-M III classification also implies a 
constraint on the mass of the donor of 
$M_{\rm d}$ = 1.2$\pm$0.2 \msun\ (Greiner \etal\ 2001a).
This proves that GRS 1915+105 is a low-mass X-ray binary (LMXB), as has been 
suggested earlier based on the IR spectral variability and on its
position on the IR H-R diagram (Castro-Tirado \etal\ 1996), though
the contrary claim has been made by  Mirabel \etal\ (1997) and 
Mart\'{\i} \etal\ (2000) that based on the detection of \ion{He}{I}, 
the donor in GRS 1915+105
should be a high-mass O or B star, and that accretion occurs predominantly
from the wind of the donor.

\subsection{The radial velocity curve}

\begin{figure}[th]
  \hspace*{-1.2cm}
      \vbox{\psfig{figure=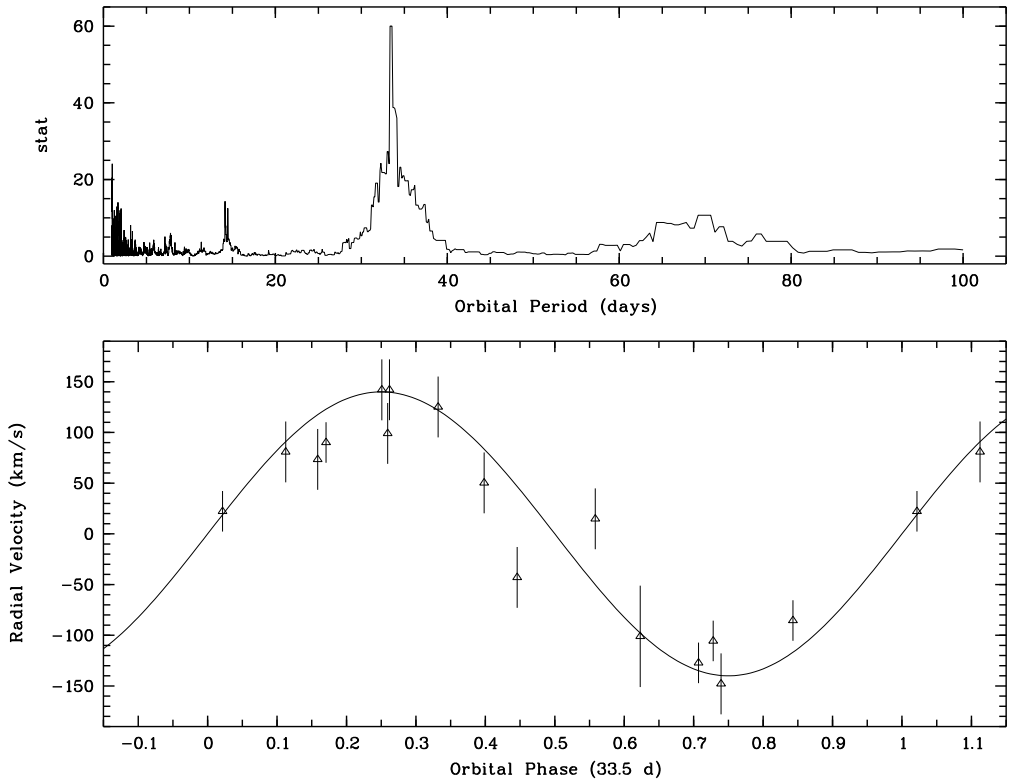,width=15.9cm,angle=270}}
    \caption[rv]{Result of the period analysis of the 
         velocity variation of the four CO bandheads.
   {\bf Top:} Scargle periodogram after heliocentric correction of the
   individual measurements.
   {\bf Bottom:} Radial velocity curve folded over the best-fit period
   of $P_{\rm orb}$ = 33.5 days. The semi-amplitude of the velocity curve
   $K_{\rm d}$ is 140$\pm$15 km/s (from Greiner \etal\ 2001b).
              }
    \label{rv}
\end{figure}

Subsequently, a series of medium-resolution spectra
in the 2.39--2.41 $\mu$m range using the VLT-Antu equipped with ISAAC
was obtained between April and August 2000 (Greiner \etal\ 2001b).
Radial velocities were measured for the individual spectra
by cross-correlation of the major CO band heads, using as template a spectrum 
of the K2\,III star HD 202135 taken with the same setting.
The results of this cross-correlation (after heliocentric correction) are 
shown in Fig. 2: 
The radial velocity periodogram shows
a clear peak at a period of 33.5 days (top panel) which is interpreted as the 
orbital period $P_{\rm orb}$ of the binary system (Greiner \etal\ 2001b). 
All measurements, folded over this period, are shown in the lower panel
of Fig. 2.
The velocity amplitude is measured to be $K_{\rm d}$ = 140$\pm$15 km/s.
With these values, the mass function, i.e. the observational
lower limit to the mass of the compact object is (Greiner \etal\ 2001b)

\begin{equation}
  f(M) \equiv {(M_{\rm c} \, {\rm sin}\, i)^3 \over (M_{\rm c} + M_{\rm d})^2} 
            = {P_{\rm orb} \, K_{\rm d}^3 \over 2 \, \pi \, G}
            = 9.5 \pm 3.0 \, M_{\odot} .
\end{equation}

\begin{table}
\caption{Spectroscopic Orbital Parameters of GRS 1915+105}
\vspace{-0.1cm}
\label{res}
\begin{center}
\begin{tabular}{ll}
      \tableline
      \noalign{\smallskip}
      Parameter     & Result \\
      \noalign{\smallskip}
      \hline
      \noalign{\smallskip}
   $T_0$ (UT)$^{(1)}$           & 2000 May 02 00:00 \\
   $T_0$ (Heliocentric)$^{(1)}$~~~~ & HJD 2\,451\,666.5$\pm$1.5~~~ \\
   $\gamma$ (km/s)              & --3$\pm$10 \\
   $K_{\rm d}$ (km/s)           & 140$\pm$15 \\
   $P_{\rm orb}$ (days)         & 33.5$\pm$1.5 \\
   $f(M)$ (\msun)               & 9.5$\pm$3.0 \\
   $M_{\rm d}$ (\msun)          & 1.2$\pm$0.2 \\
   $M_{\rm c}$ (\msun)$^{(2)}$  & 14$\pm$4 \\
 \noalign{\smallskip}
 \hline
 \end{tabular}
 \end{center}
 \vspace{-0.2cm}

 \noindent{$^{(1)}$ Time of blue-to-red crossing.
   $^{(2)}$ Using an inclination angle of $i$ = 70\deg$\pm$2\deg\ 
   (Mirabel \& Rodriguez 1994, Fender \etal\ 1999). 
  }
\end{table}

\begin{figure}[th]
    \vbox{\psfig{figure=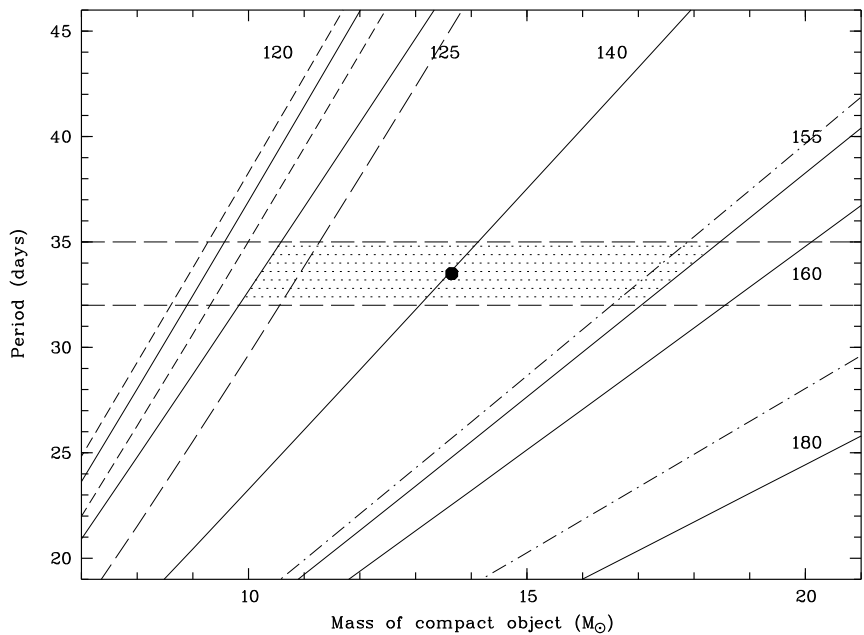,width=13.cm,angle=270}}
   \vspace{-0.3cm}
    \caption[mass]{Constraints on the black hole mass in GRS 1915+105.
 The relation of orbital period versus mass of the accreting compact object is
 plotted for various velocity amplitudes $K$ (solid lines, in km/s), assuming
  $i$=70\degs\ and a mass of the donor of 1.2 \msun.
 The dotted region shows the parameter space as constrained by the uncertainty
 of the period  and radial velocity amplitude (125--155 km/s).
 The uncertainty in the mass of the donor is shown for $K = 120$ km/s where
 the slanted dashed lines correspond to 1.0, 1.4 and 2.5 \msun, respectively
 (from left to right).
 While the formal uncertainty in the orbital inclination is only 2\degs,
 we show the effect of relaxing the assumption of the jet being
 perpendicular to the orbital plane by showing for the $K = 160$ km/s case
 the corresponding curves using $i = 79$\degs\ (at which angle eclipses
 would set in; left dash-dot curve) and $i = 61$\degs\ (right dash-dot curve).
\label{mass}}
\end{figure}

GRS 1915+105 has shown several major jet ejection events for which the
brightness and the velocities of both the approaching and the receding
blobs could be measured
(Mirabel \& Rodriguez 1994, Fender \etal\ 1999). 
From this, the angle of the jet axis relative to the 
line of sight has been determined to be $\approx$ 70\degs\ $\pm$ 2\degs\
and constant over several years. With the assumption that the
jet is perpendicular to the accretion disk and orbital plane,
this provides a reasonably accurate measure of the inclination angle $i$
of the binary system. Knowing the inclination $i$, Eq. 1 can be solved
for the mass of the accreting compact object (Fig. \ref{mass}), giving 
$M_{\rm c} = 14 \pm 4$ \msun. 
Table \ref{res} summarizes the orbital parameters of GRS 1915+105. 
Even when accounting for the 
relatively large error (which is dominated by the error in the 
determination of the velocity amplitude $K_{\rm d}$), GRS 1915+105
is the galactic binary with the largest mass function and the largest mass
of its compact object. Previous record holders were 
V404 Cyg ($f(M) = 6.07 \pm 0.05$, $M_{\rm c} = 7 - 10$ \msun;
Shahbaz \etal\ 1996)
and XTE J1118+480 ($f(M) = 6.00 \pm 0.36$, $M_{\rm c} = 6.5 - 10$ \msun;
McClintock \etal\ 2001).

We note that given the high mass-loss of the donor (which is needed
to explain the large X-ray luminosity), it is most certainly
less luminous than a non-interacting star of the same spectral type
(Langer \etal\ 2000). This
implies a larger mass for the donor, thus making the above black hole mass
a lower limit. While it is difficult to assess the exact impact of the 
mass loss on the
donor mass, the very extreme and conservative case of doubling the mass
is plotted in Fig. 3. This indicates that the black hole mass could
be higher by 1--3 \msun\ at maximum.

Distortions of the radial velocity
curve due to X-ray heating (e.g. Phillips \etal\ 1999) are likely
to be unimportant because of the long orbital period in GRS 1915+105.
Using their eq. 2, one finds that the maximum possible deviation for
GRS 1915+105 would be 8\%, as compared to their 14\% for GRO J1655-40. 
In analysing their data, Phillips \etal\ (1999) finally find that the true
velocity amplitude in GRO J1655-40 is smaller than the measured one by 
4.5\%. Thus, one could expect a 2--3\% effect in GRS 1915+105,
compared to our presently given error which amounts to 11\%.
However, since the infrared flux contribution from the secondary star
in GRS 1915+105 is so small, there is the
possibility that phase-dependent changes in the continuum near the
absorption features may result in an additional source of systematic error
in the measured value of the velocity amplitude $K_{\rm d}$.

The systemic velocity of the GRS 1915+105 binary system is   
$\gamma$ = --3$\pm$10 km/s which implies
that based on the galactic rotation curve (Fich \etal\ 1989) the kinematic
distance of GRS 1915+105 is $d = 12.1 \pm 0.8$ kpc, intermediate
between earlier estimates (Mirabel \& Rodriguez 1994, Fender \etal\ 1999).

\section{Discussion}

\subsection{Formation of a 14 \msun\ black hole}

The knowledge of the mass of the black hole in GRS 1915+105
has several implications for both the understanding of the physics
in microquasars (for a review see Mirabel \& Rodriguez 1998; Greiner 2000)
as well as some broader astrophysical concepts.
Most importantly, the formation of a 14 \msun\ black hole 
in a low-mass binary poses an interesting challenge for binary evolution.
Stellar evolution of stars in a binary system proceeds differently from 
single star evolution due to primarily the mass transfer between the system 
components and/or common-envelope phases.
There are, in general, two different paths for the black hole formation
in a binary system. First, the progenitor system could be wide, and 
during the common envelope phase the low-mass (main sequence) star of 
$\sim$1 \msun\ will spiral into the envelope of the massive giant
(progenitor of the black hole), causing the orbit to shrink 
(Kalogera 1999, 2001).
Based on the measured system parameters (Tab. 1), the deduced orbital 
separation of the binary components in GRS 1915+105 is $a$ = 108$\pm$4 \rsun.
Thus, 
orbital contraction through a common-envelope phase 
caused by the expansion of the massive progenitor to typically \gax 1000 \rsun,
is conceivable for GRS 1915+105.
Second, the evolution could start with a progenitor system smaller
than today, provided that the binary component interaction is delayed until
after He burning has ceased (Brown \etal\ 1999). In this case, the time 
between the
wind phase and the core-collapse is short, and black hole masses in the
5--10 \msun\ range have been reproduced for cases where the initial He-star
progenitor is in the 10--25 \msun\ range,
corresponding to initial primaries in the 25--45 \msun\ range 
(Brown \etal\ 2001, Kalogera 2001).
How much mass is finally lost
depends on the radii evolution of the two progenitor stars,
and it remains to be shown whether black hole masses above 10 \msun\ can
be achieved.

In order to produce BH masses \gax 10 \msun, the progenitor might have been 
a massive Wolf-Rayet star. However, Wolf-Rayet stars are known to have a much 
larger wind-loss rate, and it is therefore unclear whether higher progenitor
masses indeed will lead to higher final black hole masses. Also,
the effects of rapid rotation on the mass loss become more serious.
Even if one assumes no 
mass-loss at all from the helium star, the predicted masses for helium cores 
of massive stars can not account for BH masses \gax 12--14 \msun\ 
(Wellstein \& Langer 1999, Hurley \etal\ 2000). 
An alternative possibility to produce high-mass (\gax 10 \msun) black holes 
may be to invoke hierarchical triples as progenitors 
(Eggleton \& Verbunt 1986). 

Whatever the formation path may be, it seems that the mass of the black hole
in the binary GRS 1915+105 may provide the impetus to study the formation
of black holes with masses above 10 \msun\ in more detail.

\subsection{On the luminosity and X-ray variability of GRS 1915+105}

With the values of Table 1, the implied Roche lobe size of
the donor star is 21$\pm$4 \rsun, in good agreement with the size of
a K-M giant which therefore can be expected to fill its Roche lobe.
Thus, accretion in GRS 1915+105 is likely to occur via Roche lobe overflow.

The knowledge of the mass of the accretor in  GRS 1915+105
also allows us to possibly understand 
the rapid and large-amplitude X-ray variability 
which is seen only in GRS 1915+105 (Greiner \etal\ 1996).
Combining the observed X-ray intensity with the measured distance
immediately shows that this large-amplitude X-ray variability
occurs near or even above the Eddington limit $\dot M_{\rm Edd}$.
Such high accretion rates are seemingly never reached 
by other canonical black hole transients (e.g. GRO J1655-40)
which usually operate in the 0.1-0.2 $\dot L_{\rm Edd}$ range. At these
lower rates the accretion disks are likely gas pressure dominated, 
and thus viscously and thermally stable. In contrast, 
the uniquely high \mdot/$\dot M_{\rm Edd}$ ratio in GRS 1915+105
suggests that its inner part of the accretion disk is radiation pressure 
dominated, which in turn makes the inner disk quasi-spherical and
thermally unstable. This property potentially provides the clue
for the spectacular and unique
X-ray variability in GRS 1915+105 (Greiner \etal\ 1996)
 which has not been found in any other X-ray binary. Indeed, under the 
assumptions of a corona dissipating 50\% of the energy and a jet 
carrying a luminosity-dependent fraction of energy, Janiuk \etal\ (2000)
show that the time-dependent behaviour of GRS 1915+105 can be well reproduced
with the $\alpha$-viscosity prescription in a radiation pressure
dominated region.

Also, while it is tempting to conclude that jet ejection occurs because the
black hole can not accept this copious supply of matter, it is
important to remember that jet ejection occurs also in these other sources,
at e.g. 0.2 $\dot M_{\rm Edd}$ in GRO J1655-40,
and thus near/super-Eddington accretion cannot be the determining factor
for relativistic jets.

\subsection{On the black hole spin in GRS 1915+105 and GRO J1655-40}

The knowledge of the black hole mass also allows us to place
constraints on the question of
the black hole spin in GRS 1915+105 and GRO J1655-40.
Information on the black hole spin in both sources has been deduced from two
completely different sources: (1) Accretion disks around a (prograd) 
spinning black hole extend farther down towards the black hole, and thus
allow the temperature of the disk to be higher. Since both 
GRS 1915+105 and GRO J1655-40 exhibit a thermal component in their X-ray
spectra which has an unprecedently high temperature as compared to all other 
black hole
transients (during outbursts), it has been argued that this is due to the
spin of the black holes in these two sources, while the majority of
black hole transients has non-rotating black holes (Zhang \etal\ 1997).
(2) Several black hole binaries, including GRS 1915+105 and GRO J1655-40,
show nearly-stable quasi-periodic oscillations (QPO) in their X-ray emission.
The frequencies $f$ for these QPOs are 300 Hz in GRO J1655-40 
(Remillard \etal\ 1999) and 67 Hz in GRS 1915+105 (Morgan \etal\ 1997).
Very recently, a related second stable QPO has been discovered for each of
these (Strohmayer 2001a,b).
Most of the models proposed to explain these QPO either rely or depend
on the spin of the accreting black hole.

If the black hole mass in GRS 1915+105 is indeed no larger than 
18 \msun\ (Fig. 3), 
the deduction of the black hole spin by Zhang \etal\ (1997) is
inconsistent to any of the four models on the origin of QPOs:
\begin{enumerate}
\vspace{-0.24cm}\item If associated with the Keplerian motion at the
  last stable orbit around a (non-rotating) black hole according to
  the simple relation
  $f$ (kHz)$ = 2.2 / M_{\rm BH}$ (\msun), it gives 
  a surprising agreement with the optically determined mass for GRO J1655-40
  of 7 \msun\ (Orosz \& Bailyn 1997), 
  but is off by a factor of 2 for GRS 1915+105.
\vspace{-0.22cm}\item If associated with the trapped g-mode (diskoseismic) 
 oscillations near the inner edge of the accretion disk 
 (Okazaki \etal\ 1987, Perez \etal\ 1997, Nowak \etal\ 1997)
  would give consistency for a nearly maximally spinning black hole in
  GRO J1655-40, and a non-spinning black hole in GRS 1915+105, contrary to the
  argument that the spin in both systems should be very similar because of 
  the extremely high temperature of the accretion disk (Zhang \etal\ 1997).
\vspace{-0.22cm}\item If associated with the relativistic dragging of inertial 
  frames around a spinning black hole (Cui \etal\ 1998)  which would cause the
  accretion disk to precess, the implied specific angular momentum (spin) of 
  the black hole in
  GRS 1915+105 would be $a\sim$\,0.8, thus considerably lower than
  the $a\sim$\,0.95 deduced for GRO J1655-40. This is in conflict
  with the nearly identical accretion disk temperatures for both sources
  as deduced from the X-ray spectra, which in turn requires a larger spin
  for GRS 1915+105.
\vspace{-0.22cm}\item If associated with oscillations related to a centrifugal 
 barrier in the inner part of the accretion disk (Titarchuk \etal\ 1998),
 the product of QPO frequency and black hole mass is predicted to be
 proportional to the accretion rate, implying that the accretion rate
 in GRO J1655-40 should be a factor $\sim$10 larger than in GRS 1915+105.
 This is certainly not the case, since the black hole mass in GRO J1655-40
 is about 7 \msun\ (Orosz \& Bailyn 1997).
\vspace{-0.22cm}
\end{enumerate}
A different phrasing of the above item (1) is that one cannot argue
that based on the QPO frequency difference the mass of the black hole in
GRS 1915+105 should be a factor of 4--5 larger than that in GRO J1655-40.
This relation would only hold for the assumption that the QPO 
  frequencies are correlated to the Kepler frequency. Our mass determination
shows that the QPO frequency does {\em not} scale linearly with the mass
of the black hole. As a consequence, their origin is very likely {\em not}
related to the Keplerian frequency.

Thus, none of these four models provides a satisfactory
solution if one adopts the interpretation that the high accretion disk 
temperatures are a measure of the black hole spin (Zhang \etal\ 1997).
If, on the contrary, this latter interpretation is dropped, and the spin 
becomes a free parameter, the first three models could be applicable.

It should be noted that the applicability of the applied standard
Shakura-Sunyaev disk model 
to deduce accurate accretion disk temperatures has been also questioned 
on other grounds (Sobczak \etal\ 1999, Merloni \etal\ 2000).
In addition, there also exist alternative models, so-called slim disk 
models, which can reproduce high-temperature disks also for non-rotating 
black holes (see review by Balbus \& Hawley, or a recent application by
Watarai \etal\ 2000).

\subsection{Other implications}

Besides the above issues there are a number of other implications
of the high value of the black hole mass in GRS 1915+105
which are summarized below:
(1) It proves beyond doubt (and beyond any details in nuclear matter physics
which occasionally argue even about 7 \msun\ neutron stars; see e.g. 
Negi \& Durgapal 2001) 
the existence of stellar black holes.
(2) It has been argued that the compact objects in
most black-hole transients cluster around 7 \msun\ 
(Bailyn \etal\ 1998),
contrary to the expectation (assuming that the initial mass function is
weighted toward lower mass stars)  of a monotonic distribution of
black holes biased toward lower masses. The mass of GRS 1915+105
adds to the hints (based on V404 Cyg and Cyg X-1) that there is a 
range of black hole masses in binaries, though the bias towards low
masses (3-5 \msun) still remains to be clarified.
(3) It invalidates the counter-argument against these $\sim$7 \msun\ black 
holes being neutron stars plus a self-gravitating, 5 \msun\ accretion 
disk (Kundt 1999).
An accretion
disk with a mass of $\sim$13 \msun\ seems impossible, both
because of the higher mass-transfer which would have to last many Myrs,
as well as the unstable location of the neutron star within the disk.

Whatever the solutions to the individual implications may be, 
the microquasar GRS 1915+105 continues to surprise scientists with 
unforeseen challenges,
and to be among the most exciting X-ray sources on the sky.

\smallskip
\noindent{\it Acknowledgement:}
Though I haven't had, unfortunately, any direct collaboration with 
Jan van Paradijs, I'm grateful for having known him and having 
had the chance to interact with and benefit from him. 
Thanks to the Organizing Committee and the opportunity to present 
this work.




\begin{references}



\reference Bailyn C.D., Jain R.K., Coppi P., Orosz J.A. 1998, ApJ 499, 367

\reference Balbus S. A., Hawley J.F.  1998, Rev. Mod. Phys. 70, 1

\reference Belloni T., Mendez M., King A.R., van der Klis M.,
  van Paradijs J., 1997, ApJ 488, L109

\reference Brown G.E., Lee C.-H., Bethe H.A. 1999,
  New Astr. 4, 313

\reference Brown G.E., Lee C.-H., Tauris T.M. 2001, New Astr. 6, 331

\reference Castro-Tirado A.J., Brandt S., Lund N., \etal\
1994, ApJS 92, 469

\reference Castro-Tirado A.J., Geballe T.R., Lund N., 1996, ApJ 461, L99

\reference Cui W., Zhang S.N., Chen W. 1998, ApJ 492, L53


\reference Eggleton P.P., Verbunt F. 1986, MNRAS 220, 13p


\reference Fender R.P., Garrington S.T., McKay D.J., \etal\
1999, MNRAS 304, 865

\reference Fich M., Blitz L., Stark A.A. 1989,
 ApJ 342, 272

\reference Greiner J., Morgan E.H., Remillard R.A. 1996, ApJ 473, L107

\reference Greiner J. 2000, in Cosmic Explosions, Proc. 10th Ann.
 Astrophys. Conf. in Maryland, eds. S. Holt, W.W. Zhang, AIP 522, p. 307

\reference Greiner J., Cuby J.G., McCaughrean M.J., Castro-Tirado A.J., 
  Mennickent R.E. 2001a, A\&A 373, L37

\reference Greiner J., Cuby J.G., McCaughrean M.J. 2001b, Nat. (in press)


\reference Hurley J.R., Pols O.R., Tout C.A. 2000, MNRAS 315, 543

\reference Janiuk A., Czerny B., Siemiginowska A., 2000, ApJ 542, L33

\reference Kalogera V. 1999, ApJ 521, 723

\reference Kalogera V. 2001, in Evolution of Binary and Multipe
 Star Systems, eds. P. Podsiadlowski \etal, ASP Conf. Ser. 
 (in press; astroph/0012064)

\reference Kundt W., 1999,
F. Giovannelli \& L. Sabau-Graziati (eds.), Mem. Societa Astron. Italiana 70,
  Nos. 3/4, p. 1105

\reference Langer N., Deutschmann A., Wellstein S., H\"oflich P., 2000,
  A\&A 362, 1046

\reference Mart\'{\i} J., Mirabel I.F., Chaty S., Rodriguez L.F. 2000, 
  A\&A 356, 943

\reference McClintock J.E., Garcia M.R., Caldwell N.,
 Falco E.E., Garnavich P.M. et al.
 2001, ApJ 551, L147

\reference Merloni A., Fabian A.C., Ross R.R. 2000, MNRAS 313, 193

\reference Mirabel I.F., Rodriguez L.F. 1994, Nat. 371, 46

\reference Mirabel I.F., Bandyopadhyay R., Charles P.A., Shahbaz T., 
  Rodriguez L.F., 1997, ApJ 477, L45

\reference Mirabel I.F., Rodriguez L.F. 1998, Nat. 392, 673

\reference Morgan E.H., Remillard R.A., Greiner J. 1997, ApJ 482, 993

\reference Negi P.S., Durgapal M.C. 2001, A\&SS 275, 299

\reference Nowak M.A., Wagoner R.V., Begelman M.C., Lehr D.E. 1997, 
   ApJ 477, L91

\reference Okazaki A.T., Kato S., Fukue J. 1987, PASJ 39, 457

\reference Orosz J.A., Bailyn C.D., 1997, ApJ 477, 876

\reference Perez C.A., Silbergleit A.S., Wagoner R.V., Lehr D.E., 1997,
 ApJ 476, 589

\reference Phillips S.N., Shahbaz T., Podsiadlowski Ph. 1999, MNRAS 304, 839

\reference Remillard R.A., Morgan E.H., McClintock J.E., Bailyn C.D.,
  Orosz J.A. 1999, ApJ 522, 397

\reference Shahbaz T., Bandyopadhyay R., Charles P.A., Naylor T. 1996,
  MNRAS 282, 977

\reference Sobczak G.J., McClintock J.E., Remillard R.E.,
Bailyn C.D., Orosz J.A. 1999, ApJ 520, 776

\reference Strohmayer, T.E. 2001a, ApJ 552, L49

\reference Strohmayer, T.E. 2001b, ApJ 554, L169

\reference Titarchuk L., Lapidus I., Muslimov A. 1998, ApJ 499, 315

\reference Watarai K., Fukue J., Takeuchi M., Mineshige S. 2000,
  Publ. Astron. Soc. Japan 52, 133

\reference Wellstein S., Langer N. 1999, A\&A 350, 148

\reference Zhang N.S., Cui W., Chen W. 1997, ApJ 482, L155

\end{references}
\end{document}